\documentstyle[epsfig,aps]{revtex}

\begin{document}

\title{Hydrodynamic theory for granular gases
\thanks{URL: http://www.cec.uchile.cl/cinetica/}}
\author{Rosa Ram\'{\i}rez${}^{1,2}$, Dino Risso${}^3$,
Rodrigo Soto${}^2$
and Patricio Cordero${}^1$ \\
${}^1$  Departamento de F\'{\i}sica,
Universidad de Chile, Santiago, Chile \\
${}^2$ CECAM, ENS-Lyon, 46 All\'ee d'Italie, 69007, France
 \\
${}^3$ Departamento de F\'{\i}sica, Universidad del 
B\'{\i}o-B\'{\i}o, 
Concepci\'on, Chile}

\maketitle

\begin{abstract} 

A granular gas subjected to a permanent injection of energy is
described by means of hydrodynamic equations derived from a
moment expansion method. The method uses as reference function
not a Maxwellian distribution $f_{\sf M}$ but a distribution
$f_0 = \Phi\,f_{\sf M}$, such that $\Phi$ adds a fourth cumulant
$\kappa$ to the velocity distribution.  The formalism is applied
to a stationary conductive case showing that the theory fits
extraordinarily well the results coming from our molecular
dynamic simulations once we determine $\kappa$ as a function of
the inelasticity of the particle-particle collisions. The shape
of $\kappa$ is independent of the size $N$ of the system.

\end{abstract}


\section{Introduction}

Granular systems  subjected to a sufficiently strong excitation
may have a fluid-like behavior~\cite{jaeger,varios}.  From the
very beginning several authors have attempted to derive
hydrodynamic equations for these systems~\cite{Campbell,1r2}. 
If the
excitation of the granular system is through a permanent
injection of energy, the fluid system may stabilize to a low
density stationary gaseous state which necessarily is a
nonequilibrium state and it usually is inhomogeneous as well. To
develop the basic features of the theory of gaseous granular
systems we restrict the analysis to the simplifying inelastic
hard sphere model (IHS)~\cite{Campbell}.

Many authors studying granular gases have put particular
attention to studying the spontaneous homogeneous cooling of a
granular system using periodic boundary
conditions~\cite{Goldh,McN}. This time dependent state is called
{\em homogeneous cooling state} (HCS) and the understanding of
its properties has been improving through many
articles~\cite{GS,HCS,Brey,SotoMR}.  A crucial breakthrough was
the realization by Goldshtein and Shapiro~\cite{GS} that the
homogeneous cooling distribution function has a scaling property
with respect to the instantaneous temperature.  Such
distribution---which we will be calling $f_{\sf HCS}$---is known
in approximate forms~\cite{Ernst,Brilliantov}.  It is known,
among other things, that its fourth cumulant $\kappa$ does not
vanish and that it has a long velocity tail.

\bigskip

A nonequilibrium inhomogeneous gaseous system, on the other
hand, is described by a distorted distribution function typically
obtained from Boltzmann's equation expanding the distribution
either in gradients of the hydrodynamic fields
(Chapmann-Enskog's method)~\cite{FK} or making a moment
expansion (Grad's method)~\cite{Grad}. For normal gases the
expansion is made about the equilibrium Maxwell's distribution.

In this article we will assume that a low density nonequilibrium
granular system has a local distribution function which can be
obtained expanding about a distribution $f_0$ resembling a
$f_{\sf HCS}$ in the sense that it has a significantly
nonvanishing fourth cumulant.  We introduce a reference function
(see Eq. (\ref{f0}) below) which is a Maxwellian distorted by a
factor which incorporates the next non trivial cumulant, a
fourth cumulant, to the distribution function and then we
undertake a perturbative {\em moment expansion} \`a la Grad
about $f_0$ to solve Boltzmann's equation.  Some authors have
done some calculations in this direction but using the gradient
expansion (Chapman-Enskog) method~\cite{ChE.fHCS}.  The point is
that, without the notion of equilibrium, we expect that the
reference state, $f_0$ in our case, should resemble more the
homogeneous cooling state than the simple Maxwellian.

\bigskip

We study a two-dimensional system of hard disks, and the moment
expansion---in dimension two---is an 8 moment expansion: the
number density $n(\vec r,t)$, the velocity field $\vec v(\vec
r,t)$, the granular temperature field $T(\vec r,t)$, the
pressure tensor $P_{ij}(\vec r,t)$ and the heat flux vector
field $\vec Q(\vec r,t)$.  The dynamic variables are not the
components of the pressure tensor $I\!\!P$ itself but the
components of the symmetric traceless part 
$p_{ij}$ where $P_{ij} = p\,\delta_{ij} + p_{ij}$ and $p$ is the
hydrostatic pressure.

As it can be seen in Grad's article~\cite{Grad} or in~\cite{cou}
the method yields {\em hydrodynamic equations} for all the
fields mentioned above. In particular, the dynamic equations for
$p_{ij}$ and $\vec Q$ take the place of what would normally be
the constitutive (transport) equations of standard
hydrodynamics. This last point means that we are not assuming
any constitutive equations whatsoever, their present
counterparts are dynamic equations.

It is well established that, in the case of the IHS model for
granular systems, Boltzmann's equation is modified in that the
restitution coefficient $r$ enters solely in the gain term of
the collision integral and it does so in two forms. First the
gain term has an overall factor $r^{-2}$ and second, the
distribution functions appearing in the gain term depend upon
the precollision velocities, and these velocities depend
on~$r$~\cite{1r2}.

When the modified Boltzmann's equation is used the stemming
hydrodynamic equations get factors that depend on the {\em
inelasticity coefficient} $q$, 
\begin{equation}
                 q = \frac{1-r}{2}
\end{equation}
($q=0$ in the perfectly elastic case) except that the
mass continuity equation and the momentum balance equation
remain unchanged since mass and momentum continue being
microscopically conserved.

In the context of Boltzmann's equation a dissipative gas
satisfies the ideal gas equation of state
\begin{equation}
     p = n\,T  \label{estado}
\end{equation}
where the granular temperature $T$ is defined in energy units as
the average kinetic energy per particle.  If we were to consider
the Boltzmann-Enskog equation, then the inelasticity coefficient
would enter through the Enskog collision factor $\chi$, and the
equation of state of a normal gas and a dissipative gas would
differ, but in the present context Eq.~(\ref{estado}) holds.

Usual moment expansion methods (as Grad's is) are appropriate to
describe bulk properties. Wall effects are not well described
unless higher order momenta are included in the expansion, which
are not trivial to handle~\cite{weiss}.
Already in normal gases hydrodynamic fields may have
discontinuities at walls. In a previous article we gave a
kinetic description of a one-dimensional granular system with
theoretical tools such that our description was correct and
precise up to the walls~\cite{Rosa} but we have not generalized
yet that type of formalism to higher dimensions, hence, in the
present article, we use moment expansions.

In consequence the formalism we are going to present suffers too
of the weakness of moment expansions: it is unreliable precisely
at the points where the boundary conditions should be imposed
forcing us to trade the boundary conditions for conditions
imposed far from the walls based on the actual behavior of the
system according to our molecular dynamic simulations.

Our moment expansion method, explained in detail in Sec. II,
uses as reference distribution function a distribution $f_0$
which differs from a Maxwellian distribution in that it has a
nonvanishing fourth cumulant, $\kappa$.

The method leads to hydrodynamic equations for low density
granular systems that depend parametrically on the inelasticity
coefficient $q$ and $\kappa$. We apply this hydrodynamic
equations to a stationary and purely conductive case as
in~\cite{brey}. The value
of the fourth cumulant $\kappa$, or better, the dependence of
the fourth cumulant on $q$ is determined directly from molecular
dynamic simulations, and it turns out to be independent of the
size $N$ of the system.  The predictions that follow from our
formalism agree very well with all our simulational data.

In Sec. \ref{grad} we briefly present the moment expansion
method, in Sec. \ref{hydro} the hydrodynamic equations that
follow are given and specialized to a purely conductive case and
finally in Sec. \ref{compara} theory and simulational results
are compared. Final comments are in Sec. \ref{final}.

\section{The moment expansion method for granular 
systems} \label{grad}

Moment expansion methods  can summarily  be described as
follows. Take a velocity distribution function $f_0(\vec r, \vec
c, t)$ which is considered to be the {\em reference function}
about which an expansion is going to be made. For normal gases
the natural choice for $f_0$ is a local Maxwellian distribution
$f_{\sf M}$ written in terms of the peculiar velocity $\vec C =
\vec c -\vec v(\vec r,t)$, where $\vec v(\vec r,t)$ is the
hydrodynamic velocity. Next a set of orthonormal polynomials on
$\vec C$, $H_a(\vec C\,)$, are built in the sense that $H_{0} =
1$ and
\begin{equation}\label{ortho}
  \int H_a(\vec C\,)\, H_b(\vec C\,)\, f_0(\vec r, \vec C, t)\, 
 d\vec C = \delta_{ab}
\end{equation}
The polynomials $H_a$ are obtained simply building a base of
orthonormal polynomials starting from $H_0=1$ and from first
degree upwards. When $f_0$ is a Maxwellian the $H_a$ are Hermite
polynomials but in general they are not.

Then a solution of the form 
\begin{equation}\label{MomExpan}
   f(\vec r, \vec C, t) = \left(1 + \sum_a H_a(\vec C)
 \,R_a(\vec r, t) \right)\, f_0(\vec r, \vec C, t)
\end{equation}
is replaced in Boltzmann's equation. The $R_a$ are the moments
of $f$ with respect to the $H_a$ and they can directly be
related to the moments  associated  to $\vec c$ (the velocity
field $\vec v$),  to $\rho\,C_iC_j$ (the kinematic pressure tensor
$P_{ij}$) and  to $\frac{1}{2}\rho\,C^2\,\vec C$ (the heat flux
vector $\vec Q$).  One could go on but we have used polynomials
only containing $C_i$, $C_i\,C_j$ and $C^2\,C_i$ as Grad did.

The following step is to derive integrability conditions
multiplying the kinetic equation consecutively by the $H_n$ and
then integrating the equation over $\vec C$. The idea is to do
this up to a given order and drop all contributions coming from
polynomials of degree higher than a chosen value (up to order 3
in our case). This gives a set of hydrodynamic equations for the
different moments.

A key point is the choice of the reference function $f_0$. The
two dimensional Maxwellian $f_{\sf M}=n\,\frac{m}{2\pi\,T}
\,\exp[-m\,C^2/(2T)]$ is privileged as
the solution describing the equilibrium state of a normal gas.
Since in granular systems there is no such thing as equilibrium
a next best choice, seems a distorted Maxwellian
distribution~\cite{Ernst}
\begin{equation}\label{f0}
   f_{0} = \left[1 + \frac{\kappa}{2}\,\left(1 - {\cal C}^2 +
 \frac{{\cal C}^4}{8}\right) \,\right]\, f_{\sf M}
\end{equation}
where the dimensionless peculiar velocity is
\begin{equation}
 \vec {\cal C} = \sqrt{\frac{m}{T}}\,\vec C
\end{equation}
and $T$ is the granular temperature.  The coefficient $\kappa$
is the fourth cumulant of $f_0$ and it depends on the
inelasticity coefficient $q$ while the coefficients in front of
${\cal C}^2$ and ${\cal C}^4$ are derived from requiring that
$f_{0}$ is normalized and that
$\left<\frac{m}{2}C^2\right>_{f_0} = T$. The fourth cumulant
$\kappa$ in dimension two is
\begin{equation}\label{K4}
    \kappa = \frac{\left<{\cal C}^4\right> 
  - 2\left<{\cal C}^2\right>^2}{\left<{\cal C}^2\right>^2}\,,
\end{equation}
In the case of the homogeneous cooling state, recent articles
have justified explicit forms~\cite{Ernst,Brilliantov} for
$\kappa$ in the context of distribution functions like $f_0$. 
Their results are well approximated by
\begin{equation}\label{kappa}
   \kappa =  \frac{b_1 + b_2\,q}{1+b_3\,q}\,q \,,
\end{equation}
with
\begin{equation} \label{valores}
 b_1 = -2\, ,\qquad b_2 \approx 13.619\, , \qquad b_3 \approx
4.5969\,.
\end{equation}
The rational expression given in~(\ref{kappa}) is valid within
2.9\% for $q \le 0.08$, $r=1-2q = 0.84$. Notice that 
$\kappa(q=0)=0$ allowing to recover the elastic case. 

We describe quite satisfactorily nonhomogeneous, nonequilibrium
stationary states with a $\kappa$ as in (\ref{kappa}), but with
numerical factors different from those in Eq.~(\ref{valores}).
The latter values are valid only for the homogeneous cooling
state while our system is kept in a stationary inhomogeneous
state with appropriate boundary conditions.

For any  $\kappa$ the resulting distribution obtained
with the method summarized above is 
\begin{eqnarray}\label{fhcs}
f^{(\kappa)} &=& 
 \left[ 1+ \frac{p_{xx}}{p\,(2+\kappa)}({\cal C}_x^2-{\cal
C}_y^2) + 2 \frac{p_{xy}}{p\,(2+\kappa)} {\cal C}_x {\cal C}_y
\right. \nonumber \\ &&+ \left. \frac{Q_x}{2Q_0} \frac{({\cal
C}^2-4 -2 \kappa){\cal C}_x}{(2+5 \kappa - \kappa^2)}
+\frac{Q_y}{2Q_0} \frac{({\cal C}^2-4 -2 \kappa){\cal C}_y}{(2+5
\kappa - \kappa^2)} \right]\, f_{0}
\end{eqnarray}
where, $p_{yy}=p_{xx}$ and $Q_0=\sqrt{\frac{T}{m}}\,p$. The
distribution $f^{(\kappa)}$ shares with $f_0$ the first scalar
moments: density, temperature and fourth cumulant, for any value
of $q$. If $\kappa$ is chosen to be zero then, in~Eq.
(\ref{fhcs}), $f_{0} \to f_{\sf M}$ and $f^{(\kappa)}$ becomes
the usual Grad's distribution. Hence the whole method would be
the original method devised by Grad and, if $\kappa$ is chosen to
be Eq.~(\ref{kappa}) with coefficient values as in
Eq.~(\ref{valores}),
then $f_0$ would be what we are calling $f_{\sf HCS}$.

Given our ignorance regarding granular gases one could, in
principle, accept $f_{\sf M}$ or $f_{\sf HCS}$ as legitimate reference
functions to make the moment expansion. In the following
sections we compare the three formalisms (reference functions
$f_{\sf M}$, $f_{\sf HCS}$ and $f_0$, the latter with a $\kappa$
adjusted to the results) concluding that only the one based on
$f_0$ gives acceptable results for a sufficiently large range of
$q$.  In fact they are very good.

\section{The hydrodynamic equations} \label{hydro}

As already mentioned, the inelasticity coefficient $q$ enters
the kinetic equation in two different forms. It appears as a
factor $\frac{1}{r^2} = \frac{1}{(1-2q)^2}$ in the gain term and
it appears in the expression for the precollision velocities
which are part of the argument of the distribution functions
appearing in the gain term.  When the solution $f^{(\kappa)}$ is
inserted in Boltzmann's equation, $q$ enters in a still third
form, precisely through the $\kappa$ coefficient given by
Eqs.~(\ref{kappa}) and (\ref{fhcs}).  Expanding the collisonal
term of Boltzmann's equation in powers of $q$ and $\kappa$, the
moment method yields the following hydrodynamics equations for a
granular gas

\begin{eqnarray} 
 \frac{D n}{Dt} + n \nabla\cdot \vec v &=& 0\,, \label{masa} \\
 m\,n \frac{D\vec v}{Dt} - n\, \vec F + \nabla\cdot{I\!\!P} &=&
0\,,
\label{momen}\\
  n \frac{DT}{Dt}  +\nabla\cdot\vec Q +{I\!\!P}:\nabla\vec v 
&=& -\frac{2\,A(q)\,n\,T}{\tau}\,, \label{tempe} \\
 \frac{\partial p_{ij}}{\partial t} + 
\frac{\partial }{\partial x_k}  \left(v_k
p_{ij}\right) + \frac{1}{2} \left( 
\frac{\partial Q_i}{\partial x_j} +
\frac{\partial Q_j}{\partial x_i} - \delta_{ij}
\frac{\partial Q_k}{\partial x_k}\right) + &&  \nonumber   \\
p_{rj}\frac{\partial v_i}{\partial x_r} + p_{ri} 
\frac{\partial v_j}{\partial x_r} -
\delta_{ij} p_{rs} \frac{\partial v_s}{\partial x_r} +
p\left(\frac{\partial v_i}{\partial x_j} + 
\frac{\partial v_j}{\partial x_i} -
\delta_{ij}\frac{\partial v_r}{\partial x_r}\right) 
 &=& -\frac{B(q)}{\tau}\,p_{ij} \label{Bal-pij}\,,
\end{eqnarray}
\begin{eqnarray} 
\frac{\partial Q_k}{\partial t} + 
\frac{\partial }{\partial x_r}\left(v_r
Q_k\right) +
\frac{3}{2}\,\frac{\partial v_k}{\partial x_r}Q_r+\frac{1}{2}\,
\frac{\partial v_r}{\partial x_k}Q_r+  
\frac{1}{2}\,\frac{\partial v_r}{\partial x_r}Q_k + &&
  \nonumber \\
\frac{T}{m}\,\frac{\partial p_{kr}}{\partial x_r} +\frac{3\,p_{kr}}{m}
 \frac{\partial  T}{\partial x_r} -
\frac{p_{kr}}{m\,n} \frac{\partial P_{rs}}{\partial x_s} +
\frac{3\,p}{m} \frac{\partial T}{\partial x_k} 
 &=& -\frac{C(q)}{2\,\tau}\, Q_k \label{Bal-qk} \,,
\end{eqnarray}
where  $D/Dt \equiv \partial/\partial t + \vec v\cdot \nabla$,
\begin{equation} 
    \tau = \frac{1}{2\sigma p}\,\sqrt{\frac{m\,T}{\pi}}
\end{equation}
is a characteristic relaxation time, $\sigma$ is the
diameter of the particles, and the coefficients $A$,
$B$ and $C$ for the generic distribution $f^{(\kappa)}$ are
\begin{eqnarray}
A(q) & = & [q(1-q)+{\cal O}(q^{5})](1+\frac{3}{32}\kappa
+\frac{9}{4096}\kappa ^{2})\,, \nonumber \\ B(q) & = &
\frac{1+\frac{1}{2}q-\frac{3}{2}q^{2}+\frac{23}{32}\kappa
(1+\frac{q}{2})+\frac{\kappa ^{2}}{4096}+{\cal O}(q^{3})+{\cal O}(\kappa
^{3})}{(1+\frac{1}{2}\kappa )}\,, \label{ABC}
\\ C(q) & = &
\frac{1+\frac{13}{2}q-\frac{15}{2}q^{2}+\frac{\kappa
}{64}(206+1267q)-\frac{2415}{4096}\kappa ^{2}+{\cal O}(q^{3})+{\cal O}(\kappa
^{3})}{(1+\frac{5}{2}\kappa -\frac{\kappa ^{2}}{2})}\,. \nonumber
\end{eqnarray}

They depend on $q$ and $\kappa$, but $\kappa$ depends on $q$. 
At least for small $q$ the coefficients $A$, $B$ and $C$ are
positive ($q$ varies between 0 and $\frac{1}{2}$). The
coefficient $A$ in Eq.~(\ref{tempe}) determines the energy
dissipation in the system and consequently it vanishes in the
elastic limit while $B$ and $C$ tend to 1. In the previous
equations both $p_{ij}$ and $P_{ij}$ appear depending on which
of them gives a more compact  expression.

\vspace{1cm}

It is our hope that the above hydrodynamics is valid in a wide
variety of situations compatible with gaseous states and no
clustering~\cite{Goldh,HCS} but in this article we restrict our
study to a hydrostatic case.

We are going to consider the purely conductive regime, with no
external force, $\vec F=0$. The system of $N$ disks is in a
rectangular box of dimension $L_x\times L_y$, with thermal
walls, at $y=-\frac{1}{2}\,L_y$ and at $y=\frac{1}{2}\,L_y$,
both at temperature $T_0$, and periodic boundary conditions in
the $X$ direction. The case with no external force is quite
simple because the system is symmetric with respect to $y \to
-y$ and the pressure is uniform. If the system were
conservative, as a normal gas, this would be a homogeneous
system at thermal equilibrium, but since the system is
dissipative the temperature depends on the coordinate $y$ ($T$
has a minimum at the symmetry axis) and the problem is much less
trivial.

Since the system is purely conductive there is no velocity
field, $P_{xy}=0$, $Q_x=0$ and the other fields depend only on
the coordinate $y$. The set of
equations~(\ref{masa}-\ref{Bal-qk}) becomes \begin{eqnarray}
\label{campos}
\begin{array}{l l l l}
  P_{xx}  &= \displaystyle \frac{B-A}{B}\,p\, , &
  P_{yy} &= \displaystyle\frac{B+A}{B}\,p\, , \\
  & & & \\
 Q\,'_y &= \displaystyle
-4\,A\,p^2\,\sigma\,\sqrt{\frac{\pi}{m\,T(y)}}\, ,
 \qquad &
 T\,' &=\displaystyle  
 -\frac{\sigma\,B\,C}{3A+2B}\,\sqrt{\frac{\pi\,m}{T}}\,Q_y  \\
\end{array}
\end{eqnarray}
\noindent where the prime indicates derivative with respect to
$y$. Notice that because there is inelasticity the pressure
tensor is anisotropic in the sense that $P_{xx} \ne P_{yy}$. In
fact $P_{yy}-P_{xx} \sim A \sim q$ and they do not depend on the
coordinate $y$.

Next we are going to compare the implications of these equations
with our molecular dynamics results. To this end we should, in
principle, solve these hydrodynamic equations using the boundary
conditions associated to the particular simulations that we have
studied. This is not a straightforward task because, as we have
mentioned at the end of the introduction, the moment expansion
method (behind the previous hydrodynamic equations) does not
give a good description near boundaries. For example, if we
impose that a wall behaves as a stochastic wall at temperature
$T_0$, the observed field $T$ is not expected to take that value
near the wall. Later on it will be shown how we tackle this
problem.

From Eq.~(\ref{campos})  it is direct to derive that the
temperature field satisfies the equation
\begin{equation} 
  T\,T\,'' + \frac{1}{2} {T\,'}^2 = k^2\,\sigma^2\,p^2\,,
\label{eq:T} 
\end{equation}
where
\begin{equation} 
 k^2 \equiv \frac{4\pi\,A\,B\,C}{3A+2B} \,.\nonumber 
\end{equation}
Since the pressure is uniform, and because of Eq.~(\ref{estado}),
this equation can also be used as an equation for the inverse of
the density.

Before proceeding to solve the equations we adimensionalize the
problem defining a coordinate $\xi = y/L_y$, ($-\frac{1}{2} 
\le \xi \le \frac{1}{2}$) and we also define
\begin{equation} 
\begin{array}{r l r l}
  \bar n &= \displaystyle \frac{N}{L_x\,L_y}\,,  
&n(y) &=\bar n\,n^*(\xi)\, , \\
 & & & \\
 K &= \displaystyle
\frac{k}{2\sqrt2}\,\left(\frac{N\,\sigma}{L_x}\right) \, 
,& Q_y(y) &= \bar
n\,T_0\,\sqrt{\frac{T_0}{m}}\,\,Q_{y}^*(\xi)\,, \\
 & & & \\
T(y) &= T_0\,T^*(\xi)\,, &p &= \bar n T_0\,p^*\, . \end{array}
\end{equation}
Once~Eq.~(\ref{eq:T}) is solved and converted
to a solution for the dimensionless number density $n^*(\xi)$
the result is
\begin{equation}\label{soluc_n}
  4\,K\,|\xi| = \frac{1}{n^*(\xi)}\,
\sqrt{\frac{n_{\max}^*-n^*(\xi)}{n_{\max}^*}} +
\frac{1}{n_{\max}^*}\,{\sf
arctanh}\sqrt{\frac{n_{\max}^*-n^*(\xi)}{n_{\max}^*}}\,,
\end{equation}
where $n_{\max}^*$ is the value taken by $n^*(\xi)$ in the
middle of the system, $\xi=0$. In what follows it is shown that
$n_{\max}^*$ is a simple function of the parameter $K$ defined
above. The effect of the boundaries is traded in favor of the
observed values of $n_{\max}^*$.

The integral condition $\int n\,dx\,dy = N$ can  be cast as
\begin{equation} 
   \int_{0}^{1/2} n^* \,d\xi = \frac{1}{2}\,,
\end{equation}
but since $d\xi = (d\xi/dn^*)\,dn^*$ the integral condition
yields, \begin{equation}
   \frac{n_{b}^*}{n_{\max}^*} = 1 - \tanh^2 K\,,\label{n_b}
\end{equation}
where $n_{b}^*$ is the value that $n^*(\xi)$ takes at the two
boundaries, $n_{b}^* = n^*(\pm\frac{1}{2})$. 

Equation~(\ref{soluc_n})  evaluated at $\xi=\frac{1}{2}$ yields a
different condition over $n_{b}^*$ and it follows that
\begin{eqnarray} 
  n_{\max}^* &=& \frac{1}{2} + \frac{\sinh(2K)}{4K}\,. \label{n_max}
\end{eqnarray}
This is a strong result, it says that the dimensionless number
density at the center, $n_{\max}^*$, is determined by $K^2$ alone,
\begin{equation}\label{K2}
 K^2= a_{0}^2\,\frac{2ABC}{3A+2B}\,,
\end{equation}
where $a_{0}^2 = N\,\rho_A/\alpha$, $\alpha=L_x/L_y$ is the
aspect ratio of the box, and 
$\rho_A=(\pi/4)(N\sigma^2/(L_x L_y))$
is the fraction of area
occupied by the disks. Since $A$ vanishes in the elastic limit
and for small inelasticity coefficient $q$, $A\approx q$ while
$B\approx 1$ and $C\approx 1$ then
\begin{equation}
    K^2 \approx a_{0}^2\,q \approx
\frac{1}{\alpha}\,qN\,\rho_A\,.
\end{equation}
In the quasielastic limit then, the control parameter is
basically $qN\,\rho_A$. This result, for fixed area density, resembles
what we obtained in the one dimensional case~\cite{Rosa},
namely, that the relevant control parameter of the one
dimensional equations is $qN$.

The temperature is $T^*(\xi) = p^*/n^*(\xi)$ but since the
pressure is uniform one may be tempted to use $p^* =
n_{b}^*\,T^*_b$ with the value for $n_{b}^*$ already derived
from the theory and the value $T_{b}^*$ imposed in the
simulation. This would give a bad fit however because, as we
have been emphasizing, the formalism is not reliable near the
boundaries. Therefore we choose for $p^*$ the value $p^* =
T_{\sf min}^*\,n_{\max}^*$. We know $n_{\max}^*$ from
Eq.~(\ref{n_max}), and we take the value for $T_{\sf min}^*$
directly from our simulations. It may be said that $T_{\sf
min}^*$ is a parameter to adjust our results.

The dimensionless heat flux becomes,
\begin{equation} \label{Qy}
  Q_{y}^*(\xi) =
 \frac{3A+2B}{2\,B\,C}\, \frac{1}{a_0}\, 
 \sqrt{\frac{p^*}{n^*(\xi)}}\,\frac{p^*}
 {{n^*(\xi)}^2}\,\frac{dn^*}{d\xi}\,.
\end{equation}

\section{Simulation-theory comparison} \label{compara}

In this section we compare the simulational results with the
values given by three formalisms which use as reference
function: ($i$) $f_{\sf M}$, ($ii$) $f_{\sf HCS}$ and ($iii$)
$f_0$. We are calling $f_{\sf HCS}$ the function like $f_0$ but
with $\kappa$ defined with the values given in
Eq.~(\ref{valores}).

Since the formalisms differ by terms of higher order in $q$
their predictions are quite similar unless $qN$ is large enough.
Typically the Maxwellian theory and the one based on $f_{\sf HCS}$ 
are valid until about $qN=20$ and are reasonable until
$qN=40$ while the theory based on  $f_0$, with an adjusted
$\kappa$, is  valid for  values of $qN$ up to 200, and 
reasonably good until about  $qN \approx 300$.

\subsection{Simulational setup}

We have performed simulations of a two dimensional system of
$N=2300$, $N=3600$, $N=10000$, and $N=19600$ inelastic hard
disks inside a $L\times L$ box with lateral periodic boundary
conditions while the upper and lower walls are kept at granular
temperature~$T_0=1$. The area fraction covered by the disks was
chosen to be $\rho_A=0.01$ (in which case the nonideal
corrections to the equation of state are less than 2\%) while
the $qN$ dissipation parameter ranges from $qN=10$ up to
$qN=400$. In the $N=2300$ case, the smallest simulated system, the
ratio between the mean free path and the linear size of the
system (Knudsen number) is 0.065 and it is smaller in the other
cases. This value guarantees that not too close to the walls the
fluid has a hydrodynamic behavior. The wall temperature $T_0$ is
imposed sorting the velocity of the bouncing particles as if
they were coming from a heat bath at $T=T_0$.

In every simulation the system was relaxed from an initial
condition for a sufficiently long time.  After the relaxation we
measure local time averages of the main moments of the
distribution (i.e., $n$, $\vec{v}$, $T$, $p_{ij}$, $\vec{Q}$)
inside each one of a set of square cells.  Taking advantage of
the translation invariance in the $X$ direction, it is natural
to take horizontal averages getting, in this way, smooth
vertical profiles for the observed hydrodynamic fields.

\subsection{Theory and simulation comparison}

In order to compare theory and simulation some analysis is
needed because size effects are quite noticeable unless the
number $N$ of particles is above about 3600.  We use expression
(\ref{n_max}) for $n_{\sf max}^*$ as our point of contact between
theory and simulation and proceed to adjust the coefficients
$b_k$ in Eq.~(\ref{kappa}) so that $K$ takes the values that make
theory give the observed values for $n_{\sf max}^*$.

With this aim we first write $K$ as a rational expression $K =
a_0\,\sqrt{q}\,(1+a_1\,q)/(1+a_2\,q)$ and find the values of the
$a_k$ so that Eq.~(\ref{n_max}) reproduces the observed values
of $n_{\sf max}^*$. We have to adjust $a_0$ in spite of its
definition, given under Eq. (\ref{K2}), because the effective
values for the number of particles, the global density and the
aspect ratio get distorted since there is a layer near the
thermalizing walls which does not behave hydrodynamically. Once
this expression for $K$ is fixed, we invert (\ref{K2}) to obtain
values for $\kappa$ as a function of $q$. The size effect is in
$a_0$ alone and $\kappa(q)$ is approximately the same for
systems with $N$ larger than about 3600, as shown in Fig.
\ref{fig:kappa}. In this figure there is a solid line which is
Eq.(\ref{kappa}) with
\begin{equation}\label{good}
b_1 = -22.541\,, \quad b_2 = 6.19187\,,\quad b_3=79.0362\,.
\end{equation}
The discrepancies between this curve and the empirical values of
$\kappa$ are
less than 2\% for the whole range of~$q$ considered.  More in
detail, Fig.~\ref{fig:kappa} shows the behavior of $\kappa(q)$
for systems of different size. It is seen that in the case of
$N=2300$ (solid circles) $\kappa(q)$ is dependent on the
system's size, while the predictions in the cases $N=3600$,
$10000$, and $19600$ differ among themselves by less than 2\%. 
Only the smallest simulated system ($N=2300$) departs from this
otherwise universal shape.

In the following we present the results corresponding to the
$N=10000$ case.

\paragraph{The density:}\ As a first step and in order to check
the validity of the kinetic description (no clustering,
\cite{Goldh}) we have plotted the final configurational
positions of the particles (not shown here) for different values
of $qN$ up to $qN=400$, founding that clustering begins at about
$qN\approx 300$. In fact, we have detected that the number of
collisions per unit time increases dramatically shortly after
$qN=300$.

As a second step we check the validity  of Eq.~(\ref{n_max})
(which for fixed geometry depends only on $qN$) with
simulations.  The top Fig.~\ref{fig:n_max} shows both $n_{\rm
max}^*$ (growing curves with $qN$) and $n_{b}^*$ (the decreasing
curves). It is seen how well the values of $n_{\sf max}^*$
(solid circles) come from the $f_0$ distribution (there should
be no surprise as these are the fitted data), compared with the
prediction using the other two distributions. This graph shows
that $n_{\sf max}^*$ with the other two formalisms is good
only up to $qN\approx 40$.

Also at top Fig.~\ref{fig:n_max} shows the values predicted by
Eq.~(\ref{n_b}) and the observed values of $n_{b}^*$ (see
caption), and it is seen that all theoretical schemes give
values close to the simulational results in the considered range
of $qN$.  This is the only case where, for large $qN$,
predictions coming from $f_{\sf M}$ are better that those coming
from $f_0$, but we do not believe there is anything deep here
since our moment expansion method is not reliable near walls.

Figure~\ref{fig:n_max}, bottom,  compares theory and
simulational density profiles for $qN=30$ and $qN=200$. As it
has already been explained, the value $n_{\sf max}^*$ in
Eq.~(\ref{soluc_n}) is fixed by Eq.~(\ref{n_max}) and there are
no extra parameters to adjust. It is seen that theory in all
cases ($f_{\sf M}$, $f_{\sf HCS}$ and $f_0$) give good agreement
for low values of $qN$. However predictions for $qN = 200$, when
using $f_{\sf M}$ or $f_{\sf HCS}$ fail. It is amazing how large
$qN$ can be when $f_0$ is used. In the last case the parameter
$K$ is adjusted only from the knowledge of $n_{\max}^*$ and it
accurately predicts the behavior in almost all the volume.

From the figure it can be appreciated that near the walls
$(\xi=\pm0.5)$, as mentioned in the introduction, the theory
does not predict well the behavior of $n^*(\xi)$.

\paragraph{The temperature:}\ It has already been mentioned that
the temperature profiles exhibit a minimum at the center and
there is a temperature jump at the boundaries. In
Fig.~\ref{fig:Tmin} we show the simulational results for
$T_{\sf min}^*=T^*(\xi=0)$ and temperature
$T_{b}^*=T^*(\xi=\pm1/2)$. It is seen that for $qN=40$ the
temperature of the fluid by the walls is about 40\% lower than
the imposed value. This effect is due to dissipation and in 1D
it has been shown to be a ${\cal O}(qN)$ effect~\cite{Rosa}.

Because at present we have no theory to describe the temperature
jump that takes place near the thermal walls, and we know that
Grad's method does not give good results near boundaries, we
have chosen the observed value of the temperature at mid 
height, $T_{\sf min}^*$, as the value to use in the formalism.
The observed values decrease with $qN$ and, in the case
$N=10000$, we have adjusted them 
with the following expression 
\begin{eqnarray}
 \ln(T_{\sf min}) &=& -0.0163259\,qN +
5.27656\,10^{-5}\,\left(qN\right)^2 \nonumber \\
&&-1.86396\,10^{-7}\,\left(qN\right)^3
+4.241\,10^{-10}\,\left(qN\right)^4 \nonumber \\ &&
-4.23878\,10^{-13}\,\left(qN\right)^{5} \label{Tfit}
\end{eqnarray}
whose faithfulness is shown in Fig.~\ref{fig:Tmin}.

\medskip

Since the pressure is uniform the temperature profile can be
written directly as
\begin{equation}
       T^*(\xi) = \frac{T_{\sf min}^*\,n_{\rm
max}^*}{n^*(\xi)}\,.
\end{equation}

Figure~\ref{fig:Tprof} shows the simulational and theoretical
temperature profiles for some values of $qN$. The three upper
curves show the profiles for small $qN$ values ($qN=10$, 20, 30)
and it is seen that the predictions using $f_0$ (solid line)
give an excellent fit to the simulational data while for the
results obtained using $f_{\sf M}$ or $f_{\sf HCS}$ (dashed and
light lines respectively) the fit is only fair.  The lowest
curves shows $T$ profiles for larger values of $qN$ ($qN=50$,
100, 200). In this case only the formalism based on $f_0$ give
an acceptable description and the agreement is very good up to
$qN=200$. In the case of $qN=275$, not shown, there is still a
reasonable agreement when $f_0$ is used.

\paragraph{The pressure:} Now that the values of $n_{\sf max}^*$
and $T_{\sf min}^*$ are known and since the pressure is uniform
(both in theory and simulationally) and theory asserts that $p^* =
T_{\sf min}^*\,n_{\sf max}^*$, then we have the value of $p^*$.

From the value for the pressure and the set of equations 
(\ref{campos}) one gets the theoretical values for $P_{yy}$ in
terms of the pressure tensor. These values are compared with our
simulational data in Fig.~\ref{fig:pxxpyy}. Again the comparison
is  good when the formalism with $f_0$ is used and it is 
only fairly good with the other two formalisms.
 
As it has already been mentioned after Eq.~(\ref{campos}),
the two diagonal terms of the pressure tensor are not equal
because the system is anisotropic.

\paragraph{The heat current:}\ Equation~(\ref{Qy}) gives the
theoretical expression for the heat current $Q_{y}^*$.  At top
Fig.~\ref{fig:Qy} shows simulational data and theoretical
predictions for small values of $qN$. The three formalism
predict well the observed values.

At bottom in Fig.~\ref{fig:Qy} it is possible to see that for
$qN=200$ only the formalism using $f_0$ fits the observed data,
while the other two fail badly. In the case of $qN=275$ the
agreement is still reasonable within about 3\%.

\section{Final comments}\label{final}

In this article we have studied a bidimensional granular system
of $N$ inelastic hard disks (normal restitution
coefficient~$r$). The system is placed in a rectangular box and
it is kept in a stationary regime with upper and lower walls at
granular temperature $T_0$. The lateral walls are periodic.  The
quantity $qN$  has been used as control parameter, where
$q=\frac{1-r}{2}$, 
and simulations were made with different values of $N$ ranging
from $N=2300$ to $19600$. If the system were conservative it
would remain in a perfectly homogeneous state at temperature
$T_0$ with a homogeneous area density $\rho_A=0.01$.  Because
there is dissipation the dimensionless number density $n^*$ has
a maximum in the middle, $n_{\sf max}^*$, and the dimensionless
temperature $T^*$ has a minimum, $T_{\sf min}^*$, also at the
center of the system. The pressure is uniform and $p^* = n_{\rm
max}^*T_{\sf min}^*$.

Hydrodynamic equations were derived using a moment expansion
method. This method has the fourth cumulant, $\kappa$, of the
velocity distribution as a parameter and three possible
$\kappa$'s where considered. These are: $\kappa=0$ which implies
that the reference distribution function in a Maxwellian, and
two $\kappa$'s of the form (\ref{kappa}), one with parameters
$b_k$ as in (\ref{valores}) which corresponds to using $f_{\sf
HCS}$ as the reference function and finally using the values
$b_k$, Eq.~(\ref{good}), numerically determined to ensure that
the correct values of the density at the middle of the system
come out. The last case gives our reference function
$f_0$.

The empirical rational form of $\kappa= \kappa(q)$,
Eq.~(\ref{kappa}), in the case of the successful distribution
$f_0$ turns out to be independent of the size of the system.

When this $f_0$ is used, the comparison between the predictions
and simulation results for $n_{\sf max}^*$, $p^*$, $P_{xx}^*$,
$P_{yy}^*$ against dissipation, and the density, temperature and
heat flux profiles are very good. 

The obvious conclusion is that the formalism based on $f_0$
gives an excellent description of the behavior of the
system for values of $qN$ up to 200 and it gives a reasonable
description up to $qN$ nearly 300 (slightly above  $qN\sim 300$
clustering begins), while the other formalisms fail beyond about
$qN=40$.

Even though the fourth cumulant $\kappa$ of $f_0$ is obtained to
fit the results of a particular hydrodynamic regime, we expect
that the theory with this expression for $\kappa$ is still valid
for other regimes.

\section{Acknowledgments} This work has been partly financed by
Fondecyt research grant 296-0021 (R.R.), Fondecyt research grant
1990148 (D.R.), Fondecyt research grant 197-0786 (P.C.) and by
{\em FONDAP} grant 11980002. One of us (R.S.) acknowledges a
grant from {\em MIDEPLAN}.

\newpage

\begin{figure}[htp]
\begin{center}

\caption{The fourth cumulant $\kappa$ against $q$ for systems of
different size. The empirical values of $\kappa$ are represented
by triangles for $N=19600$, empty circles for $N=10000$, empty
squares for $N=3600$, and solid circles for $N=2300$. The solid
line corresponds to our empirical fit. See text. 
\protect\label{fig:kappa}} \end{center}
\end{figure}


\begin{figure}[htp]

\caption{At top the predicted and observed values for the
density $n_{\sf max}^*$ at the center of the box ($\xi=0$) and
$n_{b}^*$
near the
boundaries ($\xi=\pm 0.5$). The solid (open) circles correspond
to the simulational values for $n_{\sf max}^*$ ($n_{b}^*$), the
light-dashed (heavy-dashed) line corresponds to the theoretical
prediction using $f_{\sf HCS}$ ($f_{\sf M}$). The solid line
corresponds to our empirical adjustment (see text).
At bottom the predicted  
and observed density profiles for two different
values of $qN$. The open (solid) circles correspond to $qN=30$
($qN=200$). The light-dashed (heavy-dashed) line corresponds to
the theoretical prediction using $f_{\sf HCS}$ ($f_{\sf M}$).
The solid line corresponds to the prediction stemming from
$f_0$. \protect\label{fig:n_max}}
\end{figure}

\begin{figure}[htp]

\caption{Temperature $T_{\sf min}^*$ at the  center of the
channel (open circles) against dissipation $qN$. The solid line
corresponds to a fit using~Eq.~(\ref{Tfit}). The solid circles
are the observed values for the temperature $T_{b}^*$ near the
boundaries\protect\label{fig:Tmin}}
\end{figure}

\begin{figure}[htp]

\caption{ From top to bottom temperature profiles corresponding
to $qN=5, \ 10,\ 30,\ 50,\ 100$ and 200. The empty circles are
the simulational results, the dashed line (light solid line)
correspond to the predictions obtained with 
$f_{\sf M}$ ($f_{\sf HCS}$). The heavy solid lines are 
the theoretical results when
$f_0$ is used. \protect\label{fig:Tprof}}
\end{figure}

\begin{figure}[htp]

\caption{Simulational and theoretical values for the components
of the pressure tensor against dissipation $qN$. At top (bottom)
the $P_{xx}$ ($P_{yy}$) component.  In light-solid line (dashed line)
the theoretical results when using $f_{\sf HCS}$ distribution
($f_{\sf M}$ distribution). The heavy-solid line is the
theoretical result when using the $f_0$ distribution.
\protect\label{fig:pxxpyy}}
\end{figure}

\begin{figure}[htp]

\caption{Simulational and theoretical values for the heat flux
profiles. At top the three curves and set of simulational data
correspond to $qN=5$ (circles), $qN=20$ (squares) and $qN=30$
(rhombus). The dashed and light-solid lines correspond to the
predictions using $f_{\sf M}$ and $f_{\sf HCS}$ respectively,
the heavy solid line corresponds to $f_0$. At bottom are the
observed values of $Q_{y}^*$ when $qN=200$ and, the theoretical
predictions when using $f_0$ (heavy solid line), $f_{\sf M}$
(dashed line) and $f_{\sf HCS}$ (light-solid line).
\protect\label{fig:Qy}}
\end{figure}

\end{document}